 \documentclass[journal]{IEEEtran}
\usepackage{amsmath}
\usepackage{amssymb}
\usepackage{enumerate}
\usepackage{theorem}
\usepackage{graphicx,color}
\usepackage{pifont}
\usepackage[noadjust]{cite}
\usepackage{citesort}
\usepackage{amsmath,epsfig}
\usepackage[ruled,vlined]{algorithm2e}
\usepackage{amssymb,color}
\usepackage{enumerate}
\usepackage{theorem}
\usepackage{pifont}
\usepackage{multirow}

\def\PM{\kern0pt^{\textrm{{\scriptsize PM}}}\kern0pt}
\def\MMAP{\kern1pt^{\textrm{{\tiny MMAP}}}\kern-1pt}

\def\MAP{^{\textrm{{\tiny MAP}}}}

\def\Ib{{\sbm{I}}\XS}

  \def\nb{{\sbm{n}}\XS}
  
  \def\pb{{\sbm{p}}\XS}
  \def\qb{{\sbm{q}}\XS}

  \def\xb{{\sbm{x}}\XS}
  \def\yb{{\sbm{y}}\XS}
  \def\zb{{\sbm{z}}\XS}

 \def\+{^\dagger}

\usepackage{afterpage}
\usepackage{textcomp}
\RequirePackage{xspace}

\def\trans{\mbox{\tiny\textsf{T}}}

\def\qed{\ifmmode\hbox{\hfill\sqb}\else{\ifhmode\unskip\fi%
\nobreak\hfil
\penalty50\hskip1em\null\nobreak\hfil$\blacksquare$
\parfillskip=0pt\finalhyphendemerits=0\endgraf}\fi}

\RequirePackage{amsmath}
\RequirePackage{xspace}
\RequirePackage{bbm}

\def\XS{\xspace}
\DeclareMathAlphabet{\mathb}{OML}{cmm}{b}{it}
\def\sbm#1{\ensuremath{\mathb{#1}}}                
\def\sbmm#1{\ensuremath{\boldsymbol{#1}}}          

\def\Ib{{\sbm{I}}\XS}

  \def\nb{{\sbm{n}}\XS}
  
  \def\pb{{\sbm{p}}\XS}
  \def\qb{{\sbm{q}}\XS}

  \def\xb{{\sbm{x}}\XS}
  \def\yb{{\sbm{y}}\XS}
  \def\zb{{\sbm{z}}\XS}

 \def\+{^\dagger}

\def\MAP{^{\kern1pt{\rm MAP}\kern-1pt}}

\usepackage{color,soul}

\def\thetab      {{\sbmm{\theta}}\XS}

\def\PM{\kern0pt^{\textrm{{\scriptsize PM}}}\kern0pt}
\def\MMAP{\kern1pt^{\textrm{{\tiny MMAP}}}\kern-1pt} 

\def\rem#1{}                    

\def\trans{\mbox{\tiny\textsf{T}}}

\newcommand{\scal}[2]{\left\langle{{#1}|{#2}}\right\rangle}

\newcommand{\RR}{\ensuremath{\mathbb R}}

\newcommand{\prox}{\ensuremath{\mathrm{prox}}}

\newcommand{\Id}{\ensuremath{\mathrm{Id}}}

\newcommand{\vect}[1]{{\boldsymbol #1}}    

\newtheorem{property}{Property}
\newtheorem{definition}{Definition}[section]
\newtheorem{example}{Example}[section]
\title{A Hamiltonian Monte Carlo Method for Non-Smooth Energy Sampling}
\graphicspath{{./Figs/}}
\author {Lotfi Chaari \textit{IEEE Member}, Jean-Yves Tourneret, \textit{IEEE Senior member}, Caroline Chaux, 
\textit{IEEE Senior member}, and Hadj Batatia, \textit{Member, IEEE}
\\
\thanks{L. Chaari, J.-Y. Tourneret and Hadj Batatia are with the University of Toulouse, IRIT - INP-ENSEEIHT (UMR 5505),
2 rue Charles Camichel, BP 7122, Toulouse Cedex 7 France.
E-mail: firstname.lastname@enseeiht.fr.}
\thanks{C. Chaux is with LATP and CNRS UMR 7353, Aix-Marseille University, 39 rue F. Joliot-Curie, 13453 Marseille Cedex 13, France.
E-mail: caroline.chaux@latp.univ-mrs.fr.}
}
\begin{document}
\maketitle
\begin{abstract}
Efficient sampling from high-dimensional distributions is a challenging issue which is encountered in many large data recovery problems 
involving Markov chain Monte Carlo schemes. In this context, sampling using Hamiltonian dynamics is one of the recent techniques 
that have been proposed to 
exploit the target distribution geometry. 
Such schemes have clearly been shown to be efficient for multi-dimensional sampling, but 
are rather adapted to the exponential families of distributions with smooth energy function. 
In this paper, we address the problem of using Hamiltonian dynamics to sample from probability distributions having non-differentiable 
energy functions such as $\ell_1$. Such distributions are being more and more used in sparse signal and image recovery applications. 
The proposed technique uses a modified leapfrog transform involving a proximal step. 
The resulting non-smooth Hamiltonian Monte Carlo (ns-HMC) method is tested and validated on a number of experiments. 
Results show its ability 
to accurately sample according to various
multivariate target distributions. The proposed technique is illustrated on synthetic examples and is applied to an image denoising problem. 

\end{abstract}


\section{Introduction}
\label{sec:intro}
Sparse signal and image recovery is a hot topic which has gained a lot of interest during the last decades, especially after the 
emergence of the compressed sensing theory~\cite{Donoho_06}. In addition, most of the recent applications such as 
remote sensing \cite{chaux2_06} and medical image
reconstruction~\cite{Chaari_magma_2014,Boubchir_13} 
generate large and intractable data volumes that have to be processed either independently or jointly. To handle such inverse problems, Bayesian techniques have 
demonstrated their usefulness especially when the model hyperparameters are difficult to fix. Such techniques generally rely on 
a maximum a posteriori (MAP) estimation built upon the signal/image likelihood and priors. Since efficient priors generally have a 
complicated form, analytical expressions of the MAP estimators are often difficult to obtain. 
For this reason, many Bayesian techniques 
resort to Markov chain Monte Carlo (MCMC) sampling techniques~\cite{Robert_04}. 
To handle large-dimensional sampling, several techniques have 
been proposed during the last decades. In addition to the random walk Metropolis Hastings (MH) algorithm~\cite{Robert_04}, 
one can mention the work in~\cite{Richard_07} about efficient high-dimensional importance 
sampling, the Metropolis-adjusted Langevin algorithm (MALA) of~\cite{Roberts_96,pereyra_13},  
or the high-dimensional Gaussian sampling methods of~\cite{Orieux_12,gilavert_14}. 
To handle log-concave smooth probability distributions, a Hamiltonian Monte Carlo (HMC) sampling technique has recently been proposed 
in~\cite{Hanson_01,neal2010mcmc}. This technique uses the analogy with the kinetic energy conservation in physics 
to design efficient proposals that better follow the target distribution's geometry. HMC has recently been investigated in a number of 
works dealing with multi-dimensional sampling problems for various applications~\cite{Pakman13,Altmann_2014}, which demonstrates the 
efficiency of such sampling schemes. Efficient sampling is obtained using these strategies where the convergence and mixing properties 
of the simulated chains are improved compared to classical sampling schemes such as the Gibbs and MH algorithms. 
However, these techniques are only adapted to log-concave probability distributions with smooth energy functions since the gradient of these 
functions has to be calculated. This constraint represents a real limitation in some applications 
where sparsity is one of the main processing ingredients especially 
for large data. Indeed, sparsity promoting 
probability distributions generally have a non-differentiable energy function such as the Laplace or the 
generalized Gaussian (GG)~\cite{Wai_95} 
distributions which involve $\ell_1$ and $\ell_p$ energy functions, respectively. Such distributions have been used as priors for the 
target signals or images in a number of works where inverse problems are handled in a Bayesian 
framework~\cite{Simoncelli_96,Moulin_98,Chaari_TSP_2010}.
Using the HMC technique in this context is therefore not possible. \\
This paper presents a modified HMC scheme that makes it possible to sample from log-concave probability distributions with non-differentiable
energy functions. 
The so called non-smooth HMC (ns-HMC) sampling scheme relies on a modified leapfrog transform~\cite{Hanson_01,neal2010mcmc} 
that circumvents 
the non-differentiability of the target energy function. The modified leapfrog transform relies on the sub-differential and proximity 
operator concepts~\cite{Bauschke_combettes2011}. 
The proposed scheme is validated on a sampling example where samples are drawn from a 
GG distribution with different shape parameters. It is also illustrated on a signal recovery problem 
where a sparse regularization scheme is applied to recover a high-dimensional signal. 

The rest of the paper is organized as follows. Section~\ref{sec:pb} formulates the problem of non-smooth sampling for large data using 
Hamiltonian dynamics. 
Section~\ref{sec:sampling} presents the  
proposed ns-HMC sampling scheme. This technique is then validated in Section ~\ref{sec:simu} to illustrate its efficiency for sampling from 
non-smooth log-concave distributions. 
Finally, some conclusions and perspectives are drawn in Section~\ref{sec:cl}.

\section{Problem formulation}\label{sec:pb}
Let us consider a signal of interest $\xb \in \RR^N$  and let $f(\xb;\thetab)$ be its probability distribution function which is parametrized 
by the parameter vector $\thetab$. In this work, we focus on an exponential family of distributions such that
\begin{equation}\label{eq:expon}
 f(\xb ; \thetab) \propto \exp [ - E_\thetab(\xb)]
\end{equation}
where $E_\thetab(\xb)$ is the energy function. As stated above, in this paper we concentrate on sampling from
the class of log-concave probability densities, which means that the energy function $E_\thetab$ 
is assumed to be convex but not necessarily differentiable. In addition, we will also make the assumption that 
$E_\theta$ belongs to $\Gamma_0(\RR)$, the class of proper lower semi-continuous convex functions from $\RR$ to $]-\infty,+\infty]$. Finally, 
we will consider probability distributions from which direct sampling is not possible and requires the use of an 
acceptance-rejection step. Example~\ref{ex:gg} presents the case of the GG distribution which satisfies the above mentioned assumptions.



\begin{example}\label{ex:gg}
Let $\gamma >0$ and $p \geq 1$ two real-positive scalars. The generalized Gaussian distribution $\mathrm{GG}(x;\gamma,p)$ is defined 
by the following probability density function 
  \begin{equation}\label{eq:gg}
\mathrm{GG}(x;\gamma,p)  = \frac{p}{2 \gamma^{1/p} \Gamma(1/p)} \exp{\left(-\frac{|x|^p}{\gamma}\right)}
\end{equation}
for $x\in \RR$.
\end{example}
Except for particular values of $p$ such as $p=2,4,\ldots$, the energy function 
$E_\thetab(x) = \frac{|x|^p}{\gamma}$ is not differentiable (where $\thetab= (\gamma,p)$). In what follows, we are interested in efficiently drawing samples according to the probability distribution $f$ defined 
in \eqref{eq:expon}. The following section describes the 
proposed non-smooth sampling scheme.

\section{Non-smooth sampling}\label{sec:sampling}

\subsection{Hamiltonian Monte Carlo methods}
\label{sec:HMCmethods}
HMC methods~\cite{Hanson_01,neal2010mcmc,Altmann_2014} are powerful tools that use 
the principle of Hamiltonian dynamics. These methods have been originally proposed by analogy to 
kinetic energy evolution. They usually provide better mixing and convergence properties than standard schemes. 
Let us assume that $E_\thetab(\xb)$ represents a \emph{potential} energy function.
If we associate to $\xb$ a \emph{momentum} vector $\qb \in \RR^N$, the energy function $H_\thetab$ for the Hamiltonian dynamics is
the combination of the potential energy $E_\thetab(\xb)$ and a \emph{kinetic} energy $K(\qb)$, i.e.,

\begin{equation}\label{eq:ham}
 H_\thetab(\xb,\qb) = E_\thetab(\xb) + K(\qb).
\end{equation}
The energy function $ H_\thetab$ is called the \emph{Hamiltonian} and completely describes the considered system. 
For simplicity reasons, a quadratic kinetic energy corresponding to a unitary diagonal covariance matrix is usually assumed so that
$K(\qb) = \frac{1 }{2}\qb^{\trans}\qb$.  
The Hamiltonian's motion equations determine the evolution of the state as a function of time $t$~\cite{neal2010mcmc}
\begin{align}\label{eq:hammotion}
 \frac{dq}{dt} = & \frac{\partial H_\thetab}{\partial x} \nonumber \\
 \frac{dx}{dt} = & -\frac{\partial H_\thetab}{\partial q}
\end{align} where $\partial$ denotes the partial derivative operator. 
These equations define a transformation $\mathcal{F}_s$ that maps the state of the system at time $t$ to the state at time $t+s$.
The distribution of the Hamiltonian dynamics energy defined in \eqref{eq:ham} is therefore given by

\begin{align}\label{eq:hamdist}
 f_\thetab(\xb,\qb) &\propto \exp \left[  - H_\thetab(\xb,\qb) \right] \nonumber \\
 &\propto f(\xb;\thetab) \exp\left( - \frac{\qb^{\trans} \qb}{2} \right).
\end{align}

HMC methods iteratively proceed by alternate updates of samples $\xb$ and $\qb$ drawn according to the distribution \eqref{eq:hamdist}. 
At iteration $\# r$, the HMC algorithm starts with the current values of vectors $\xb^{(r)}$ and $\qb^{(r)}$.
Two steps have then to be performed. The first one proceeds by an update of the momentum vector leading to $\bar{\qb}^{(r)}$
by sampling according to the multivariate Gaussian distribution
$\mathcal{N}(\textbf{0}, \Ib_N)$, where $\Ib_N$ is the $N \times N$ identity matrix.
The second step updates both momentum $\qb$ and position $\xb$ by proposing two candidates $\xb^{*}$ and $\qb^{*}$.
These two candidates are generated by simulating the Hamiltonian dynamics, which are discretized using some discretization techniques such
as the Euler or leapfrog methods. For instance, the discretization can be performed using $L_f$ steps of the leapfrog method
with a stepsize $\epsilon > 0$.  The $l$th leapfrog discretization will be denoted by $T_s$ and can be summarized as follows
\begin{align}
 \qb^{(r,(l+\frac{1}{2}) \epsilon)} = &  \qb^{(r,l \epsilon)}  - \frac{\epsilon}{2} \frac{\partial E_\thetab}{\partial \xb^{\trans}} \left(\xb^{(r,l\epsilon)}  \right) \label{eq:leapfrog1}\\
 \xb^{(r,(l+1) \epsilon)} = &  \xb^{(r,l \epsilon)}  + \epsilon  \qb^{(r,(l+\frac{1}{2}) \epsilon)} \label{eq:leapfrog2}\\
 \qb^{(r,(l+1) \epsilon)} = &  \qb^{(r,(l+\frac{1}{2}) \epsilon)}  - \frac{\epsilon}{2} \frac{\partial E_\thetab}{\partial \xb^{\trans}} \left(\xb^{(r,(l+1) \epsilon)} \right)\label{eq:leapfrog3}.
\end{align}
After the $L_f$ steps, the proposed candidates are given by $\qb^{*} = \qb^{(r,\epsilon L_f)}$ and $\xb^{*} = \xb^{(r,\epsilon L_f)}$.
These candidates are then
accepted using the standard MH rule, i.e., with the following probability
\begin{equation}
\label{eq:MHruleb}
\min \bigg\{ 1,\exp\left[ H_\thetab(\xb^{(r)},\bar{\qb}^{(r)}) - H_\thetab(\xb^*,\qb^*)\right] \bigg\}
\end{equation}
where $H_\thetab$ is the energy function defined in \eqref{eq:ham}.

\subsection{Non-smooth Hamiltonian Monte Carlo schemes}
\label{sec:proxHMC}
The key step in standard HMC sampling schemes is the approximation of the Hamiltonian dynamics. 
This approximation allows the random simulation of uncorrelated samples according to a target distribution while exploiting the
geometry of its corresponding energy. In this section, we propose two non-smooth Hamiltonian Monte Carlo (ns-HMC) schemes 
to perform this approximation for
non-smooth energy functions. The first scheme is based on the subdifferential operator while the second one is based on proximity operators. 
For both schemes, the whole algorithm to sample $\xb$ and $\qb$ is detailed in Algorithms 1 and 2. These algorithms describe all the necessary 
steps to sample from a log-concave target distribution.\\

\subsubsection{Scheme 1 - subdifferential based approach}\hfill\\
Let us first give the following definition of the sub-differential and a useful example.
\begin{definition}\cite[p. 223]{Bauschke_combettes2011}
 Let $\varphi$ be in $\Gamma_0(\RR)$, the class of lower semi-continuous convex functions from $\RR$ to $]-\infty,+\infty]$.
 The sub-differential of $\varphi$ is the set
 $\partial \varphi (x) = \{\rho \in \RR | \; \varphi(\eta) \geq \varphi(x) + \scal{\rho}{\eta-x} \forall \eta \in \RR \}$, 
 where $\scal{\cdot}{\cdot}$ defines the standard scalar product.
 Every element $\rho \in \partial \varphi (x)$ is a sub-gradient of $\varphi$ at point $x$. If $\varphi$ is differentiable, the
 sub-differential reduces to its gradient: $\partial \varphi(x) = \{ \nabla \varphi (x) \}$.
\end{definition}

\begin{example}\label{ex:l1}
Let $\varphi$ be defined as
\begin{align}
 \varphi: \; \RR &\longmapsto \RR \nonumber \\
 x &\longrightarrow |x|.
\end{align}
The sub-differential of $\varphi$ at $x$ is defined by
\begin{equation}
\partial \varphi (x) =
\begin{cases}
\{\mathrm{sign}(x)\} & \mbox{if $x\neq 0$}\\
[-1,1] & \mbox{if $x= 0$}.
\end{cases}
\end{equation}
If in addition we consider a scalar $\lambda \in \RR_+$ and we call $\varphi_\lambda(\cdot) = \lambda \varphi(\cdot)$, then we have 
$\partial \varphi_\lambda (x) = \lambda \partial \varphi (x)$ for every $x \in \RR$ \cite[Prop. 16.5]{Bauschke_combettes2011}.
\end{example}

For distributions with smooth energy, we propose to use the leapfrog method whose basic form requires to compute the
gradient of the potential energy $E_\thetab(\xb)$. Since we cannot determine this gradient for non-smooth energy functions, 
we resort to the following reformulation of the
leapfrog scheme by using the concept of sub-differential introduced hereabove
\begin{align}
 \qb^{(r,(l+\frac{1}{2}) \epsilon)} = &  \qb^{(r,l \epsilon)}  -
 \frac{\epsilon}{2} \rho \left(\xb^{(r,l\epsilon)}\right) \label{eq:leapfrog1bb} \\
 \xb^{(r,(l+1) \epsilon)} = &  \xb^{(r,l \epsilon)}  + \epsilon  \qb^{(r,(l+\frac{1}{2}) \epsilon)} \label{eq:leapfrog2bb}\\
 \qb^{(r,(l+1) \epsilon)} = &  \qb^{(r,(l+\frac{1}{2}) \epsilon)}  -
  \frac{\epsilon}{2} \rho \left(\xb^{(r,(l+1) \epsilon)}\right) \label{eq:leapfrog3bb}
\end{align}
where $\rho \in \partial E_\thetab $ is sampled uniformly in the sub-differential of $E_\thetab$. This discretization scheme will be denoted by 
$T'_{s}$. If $E_\thetab(\xb)$ is differentiable, 
the mapping $T'_{s}$ in \eqref{eq:leapfrog1bb}, \eqref{eq:leapfrog2bb} and \eqref{eq:leapfrog3bb}
exactly matches the conventional HMC mapping $T_{s}$ in ~\eqref{eq:leapfrog1}, \eqref{eq:leapfrog2} and \eqref{eq:leapfrog3}.\\

As for the standard HMC scheme, the proposed candidates are defined by $\qb^{*} = \qb^{(r,\epsilon L_f)}$
and $\xb^{*} = \xb^{(r,\epsilon L_f)}$ that can be computed after $L_f$ leapfrog steps.
These candidates are then
accepted based on the standard MH rule defined in \eqref{eq:MHruleb}. 
The resulting sampling algorithm is summarized in Algorithm~\ref{algo:proxMHlpsub}.

\begin{algorithm}\label{algo:proxMHlpsub}
\SetAlgoLined
- Initialize with some $\vect{x}^{(0)}$.\\
- Set the iteration number $r=0$, $L_f$ and $\epsilon$\;
\For{$r=1\ldots S$}{
- Sample $\qb^{(r,0)}\sim \mathcal{N}(\textbf{0},\Ib_N)$\;
- Compute $\qb^{(r,\frac{1}{2} \epsilon)} =  \qb^{(r,0)}  -
 \frac{\epsilon}{2} \rho(\xb^{(r,0)} ) $\; 
 - Compute $\xb^{(r,\epsilon)} =  \xb^{(r,0)}  + \epsilon  \qb^{(r,\frac{1}{2} \epsilon)} $\;
 \For{$l_f=1$ to $L_f-1$}{
 * Compute $\qb^{(r,(l_f+\frac{1}{2}) \epsilon)} =  \qb^{(r,l_f \epsilon)}  -
 \frac{\epsilon}{2} \rho(\xb^{(r,l_f\epsilon)})$\;
 * Compute $\xb^{(r,(l_f+1) \epsilon)} = \xb^{(r,l_f \epsilon)}  + \epsilon  \qb^{(r,(l_f+\frac{1}{2}) \epsilon)}$\;
 }
 - Compute $\qb^{(r,(L_f+\frac{1}{2}) \epsilon)} =  \qb^{(r,L_f \epsilon)}  -
 \frac{\epsilon}{2} \rho(\xb^{(r,L_f\epsilon)}) $\;
 - Apply standard MH acceptation/rejection rule by taking $\qb^{*} = \qb^{(r,\epsilon L_f)}$ and $\xb^{*} = \xb^{(r,\epsilon L_f)}$\;
}
\caption{Gibbs sampler using Hamiltonian dynamics for non-smooth log-concave probability distributions: Scheme 1.}
\end{algorithm}

Note that we do not need to account for any additional term in the acceptance ratio in \eqref{eq:MHruleb} since volume preservation is 
ensured by the Metropolis
update. Volume preservation is equivalent to having an absolute value of the Jacobian matrix determinant for the mapping $T_s$ 
equal to one, and
is due to the fact that candidates are proposed according to Hamiltonian dynamics. 
More precisely, volume preservation can be easily demonstrated by using 
the concept of Jacobian matrix approximation~\cite{Jeyakumar_98} such as the \textit{Clarke} 
generalization~\cite{Clarke_83}, and by conducting calculations
similar to~\cite[Chapter 5, p. 118]{neal2010mcmc}. 
To facilitate the reading of the paper, the volume preservation issue is addressed in Appendix~\ref{append:volume}. \\


\subsubsection{Scheme 2 - proximal based approach}\hfill \\

\noindent  Since the calculation of the sub differential is not straightforward for some classes of convex functions, 
a second scheme modifying the leapfrog steps \eqref{eq:leapfrog1bb}, \eqref{eq:leapfrog2bb} and \eqref{eq:leapfrog3bb}  
can be considered by using the concept of \emph{proximity operators}. 
These operators have been found to be fundamental in a number of
recent works in convex optimization \cite{Combettes_05,Chaux_C_07,Chaari_MEDIA_2011}, and more 
recently in~\cite{Atchade_14} where stochastic proximal algorithms have been investigated. Let us first recall the following definition. 

\begin{definition}\cite[Definition 12.23]{Bauschke_combettes2011}\cite{Moreau_65}
 Let $\varphi \in \Gamma_0(\RR)$. For every $x \in \RR$, the function $\varphi+\Vert .-x \Vert^2/2$
 reaches its infimum at a unique point referred to as proximity operator and denoted by $\mathrm{prox}_{\varphi}(x)$.
\end{definition}

\begin{example}
For the function $\varphi$ defined in Example~\ref{ex:l1}, the proximity operator is given by
\begin{equation}
\mathrm{prox}_{\varphi}(x) = \mathrm{sign}(x) \max\{ |x|-1,0\} \; \forall  x \in \RR.
\end{equation}
\end{example}
Many other examples and interesting properties that make this tool very powerful and commonly used in the recent optimization literature are given
in~\cite{Combettes_Pesquet_chapter11}. One of these properties in which we are interested here is given in the following property.

\begin{property}\cite[Prop. 3]{douglas_56}
\label{prp:subdiffprox}
 Let $\varphi \in \Gamma_0(\RR)$ and $ x \in \RR$. There exists a unique point $\widehat{x}\; \in \RR$ such that
 $ x- \widehat{x} \in \partial \varphi (\widehat{x})$. Using the proximity operator definition hereabove, it turns out that
 $\widehat{x} = \prox_\varphi (x)$.
\end{property}

By modifying the discretization scheme $T_s$, we propose the following $l$-th 
leapfrog discretization scheme denoted by $T''_{s}$
\begin{align}
 \qb^{(r,(l+\frac{1}{2}) \epsilon)} = &  \qb^{(r,l \epsilon)}  -
 \frac{\epsilon}{2} \left[\xb^{(r,l\epsilon)} - \prox_{E_\thetab}(\xb^{(r,l\epsilon)}) \right] \label{eq:leapfrog1b} \\
 \xb^{(r,(l+1) \epsilon)} = &  \xb^{(r,l \epsilon)}  + \epsilon  \qb^{(r,(l+\frac{1}{2}) \epsilon)} \label{eq:leapfrog2b}\\
 \qb^{(r,(l+1) \epsilon)} = &  \qb^{(r,(l+\frac{1}{2}) \epsilon)}  -
  \frac{\epsilon}{2} \times \nonumber \\
  &\left[\xb^{(r,(l+1) \epsilon)} - \prox_{E_\thetab}(\xb^{(r,(l+1) \epsilon)}) \right]\label{eq:leapfrog3b}.
\end{align}
If $E_\thetab(\xb)$ is differentiable, the mapping $T''_{s}$ in \eqref{eq:leapfrog1b}, \eqref{eq:leapfrog2b} and \eqref{eq:leapfrog3b}
exactly matches the mapping $T_{s}$ in ~\eqref{eq:leapfrog1}, \eqref{eq:leapfrog2} and \eqref{eq:leapfrog3}. The only difference 
is that the sub-differential of the mapping $T''_{s}$ is evaluated in $\prox_{E_\thetab}(\xb)$ instead of $\xb$. As for scheme~1, the proposed candidates are given by $\qb^{*} = \qb^{(r,\epsilon L_f)}$
and $\xb^{*} = \xb^{(r,\epsilon L_f)}$ after $L_f$ leapfrog steps. 
These candidates are then
accepted based on the standard MH rule \eqref{eq:MHruleb}. 

The Gibbs sampler resulting from the transformation $T''_{s}$ is summarized in Algorithm~\ref{algo:proxMHlp}. 
As well as for Algorithm~\ref{algo:proxMHlpsub}, and due to the presence of the MH acceptance rule, the elements $\xb^{(r)}$ generated by this algorithm are asymptotically distributed 
according to the target distribution $f(\xb;\thetab)$ defined in \eqref{eq:expon}.

\begin{algorithm}\label{algo:proxMHlp}
\SetAlgoLined
- Initialize with some $\vect{x}^{(0)}$.\\
- Set the iteration number $r=0$, $L_f$ and $\epsilon$\;
\For{$r=1,\ldots, S$}{


- Sample $\qb^{(r,0)}\sim \mathcal{N}(\textbf{0},\Ib_N)$\;
- Compute $\qb^{(r,\frac{1}{2} \epsilon)} =  \qb^{(r,0)}  -
 \frac{\epsilon}{2} \left[\xb^{(r,0)} - \prox_{E_{\thetab}}(\xb^{(r,0)} )\right] $\;
 - Compute $\xb^{(r,\epsilon)} =  \xb^{(r,0)}  + \epsilon  \qb^{(r,\frac{1}{2} \epsilon)} $\;
 \For{$l_f=1$ to $L_f-1$}{
 * Compute $\qb^{(r,(l_f+\frac{1}{2}) \epsilon)} =  \qb^{(r,l_f \epsilon)}  -
 \frac{\epsilon}{2} \left[\xb^{(r,l_f\epsilon)} - \prox_{E_{\thetab}}(\xb^{(r,l_f\epsilon)}) \right]$\;
 * Compute $\xb^{(r,(l_f+1) \epsilon)} = \xb^{(r,l_f \epsilon)}  + \epsilon  \qb^{(r,(l_f+\frac{1}{2}) \epsilon)}$\;
 }
 - Compute $\qb^{(r,(L_f+\frac{1}{2}) \epsilon)} =  \qb^{(r,L_f \epsilon)}  -
 \frac{\epsilon}{2} \left[\xb^{(r,L_f\epsilon)} - \prox_{E_{\thetab}}(\xb^{(r,L_f\epsilon)}) \right]$\;
 - Apply standard MH acceptation/rejection rule by taking $\qb^{*} = \qb^{(r,\epsilon L_f)}$ and $\xb^{*} = \xb^{(r,\epsilon L_f)}$\;

}
\caption{Gibbs sampler using Hamiltonian dynamics for non-smooth log-concave probability distributions.}
\end{algorithm}

\subsubsection{Discussion}\hfill \\

\begin{figure}[!ht]

\centering
\hspace{-.3cm}\includegraphics[height=7cm,width=9cm]{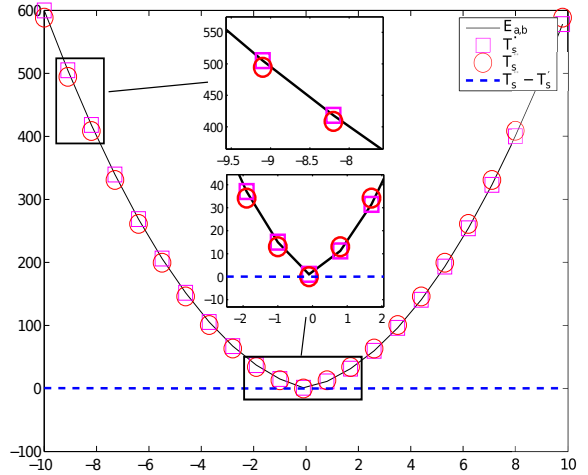}
 \caption{The potential energy $E_{a,b}$ (solid black line) in \eqref{eq:hamexp} ($a=10, b=5$) and its discretizations using the modified leapfrog schemes $T'_{s}$ (squares) 
 and $T''_{s}$ (circles), as well 
 as the difference between the 
 two discretizations $T''_{s} - T'_{s}$ (dashed blue line).}\label{fig:leapfrog}
 \end{figure}

\begin{figure}[!ht]

\centering
\hspace{-.3cm}\includegraphics[height=7cm,width=9cm]{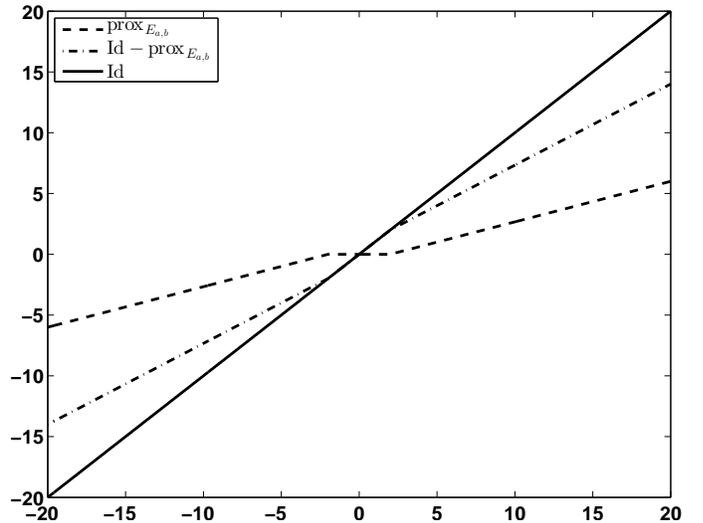}
 \caption{The proximity operator $\prox_{E_{a,b}}$, the identity function ($\mathrm{Id}$) and the difference 
 $\rm{Id} - \prox_{E_{a,b}}$ for $a=b=2$.}\label{fig:prox}
 \end{figure}

Fig.~\ref{fig:leapfrog} illustrates the use of the adopted discretization schemes associated with algorithms 1 and 2 in approximating a Hamiltonian made up of 
a quadratic kinetic energy and a potential energy having the following form
\begin{equation}\label{eq:hamexp}
 E_{a,b}(x) = a|x| + bx^2
\end{equation}
where $(a,b) \in (\mathbb{R}_+^*)^2$. For this potential energy, the sub-differential can be analytically calculated and is given by

\begin{equation}
 \partial E_{a,b} = a \partial \varphi + (2b) \rm{Id}
\end{equation}
where $\partial \varphi$ is defined in Example~\ref{ex:l1} and $\rm{Id}$ is the identity operator. 
The two proposed algorithms can therefore be compared for this 
example. 
Fig.~\ref{fig:leapfrog} shows that the discretized energy is close to the continuous one for the two mappings $T'_{s}$ and $T''_{s}$. 
Moreover, the slight difference $T''_{s} -T'_{s}$ between the two mappings shows that the two discretization schemes perform very similarly close to the critical region of non-differentiability (the interval 
$[-\varepsilon,\varepsilon]$ with small $\varepsilon \in \RR_+$, see the zoom around the origin in Fig.~\ref{fig:leapfrog}). 
Fig.~\ref{fig:prox} illustrates the shape of the proximity operator for the 
considered energy function $E_{a,b}$, as well as the identity function $\rm{Id}$ and the difference ${\rm{Id}} - \prox_{E_{a,b}}$. 
This figure 
clearly shows that, due to the thresholding property of the proximity operator, $x \simeq x - \prox_{E_{a,b}}x$ for 
$x \in [-\varepsilon,\varepsilon]$. In particular, for the considered 
example, we have $ x = x - \prox_{E_{a,b}}x$ for every $x \in [\frac{-a}{b+1},\frac{a}{b+1}]$. This comparison confirms that the 
two schemes perform similarly especially close to the non-differentiability point.

Since Algorithm 2 is more general than Algorithm 1 and allows us to handle energies for which the sub-differential is not straightforward 
(while performing well especially close to the critical regions), we will focus on this discretization scheme for our experiments.  

\section{Experimental validation}
\label{sec:simu}
This section validates the proposed ns-HMC scheme for non-smooth log-concave distributions through three experiments. 
The two first experiments consider the GG distribution whose energy function is non-differentiable for the 
values of the shape parameter considered here ($p=1$ and $p=1.5$). For the third experiment, a Laplace distribution (GG distribution with 
$p=1$) is used for an image denoising problem where the clean image is recovered from noisy measurements using a Bayesian regularization scheme involving 
a sampling technique based on the proposed ns-HMC algorithm.


\subsection{Experiment 1: 1D sampling}

\begin{figure}[!htp]
\centering
\begin{tabular}{cc}
\rotatebox{90}{\hspace{1.5cm}$p=1$, $\lambda=1$}&\includegraphics[height=4.5cm,width=6.5cm]{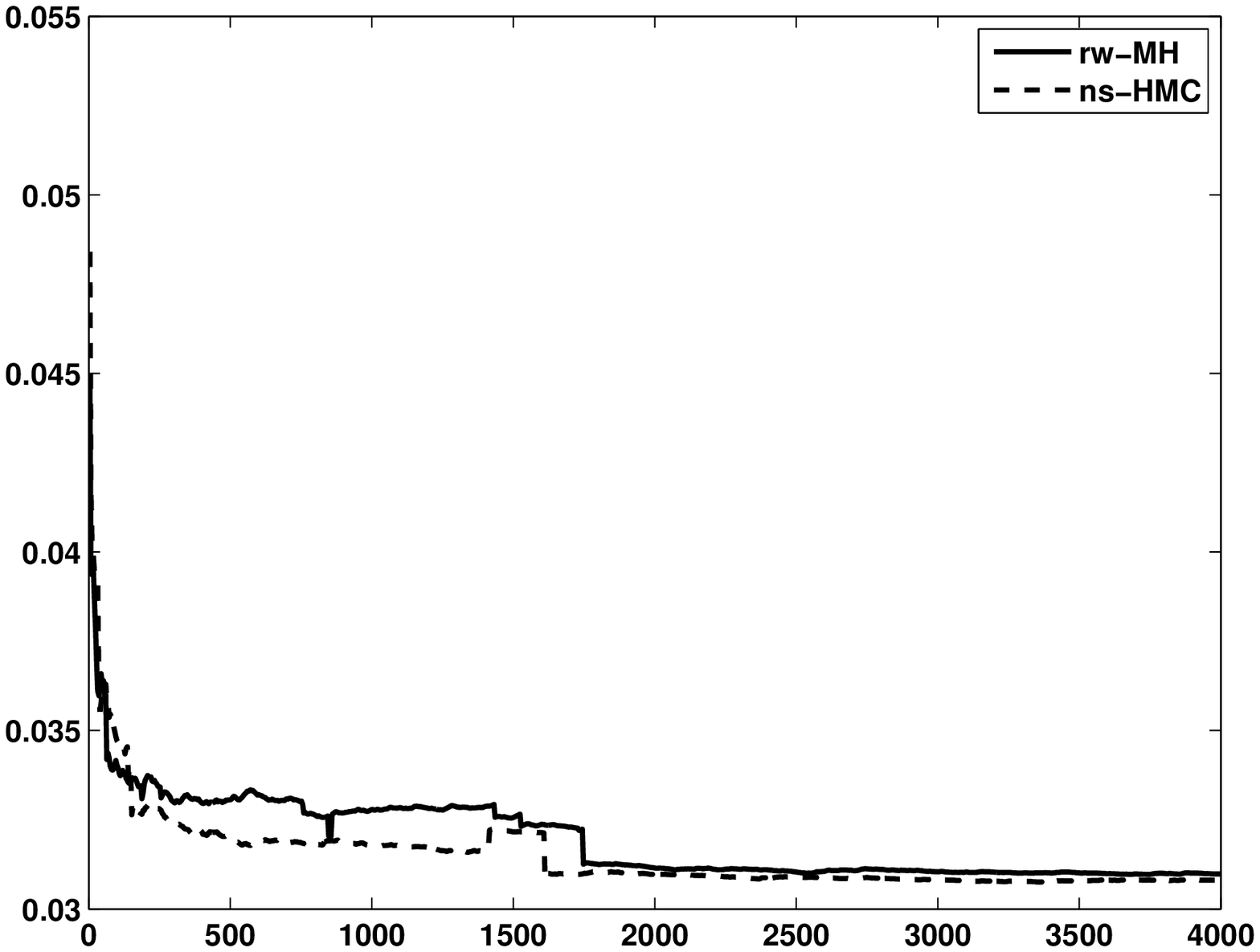}\\
\rotatebox{90}{\hspace{1.5cm}$p=1.5$, $\lambda=1$}&\includegraphics[height=4.5cm,width=6.5cm]{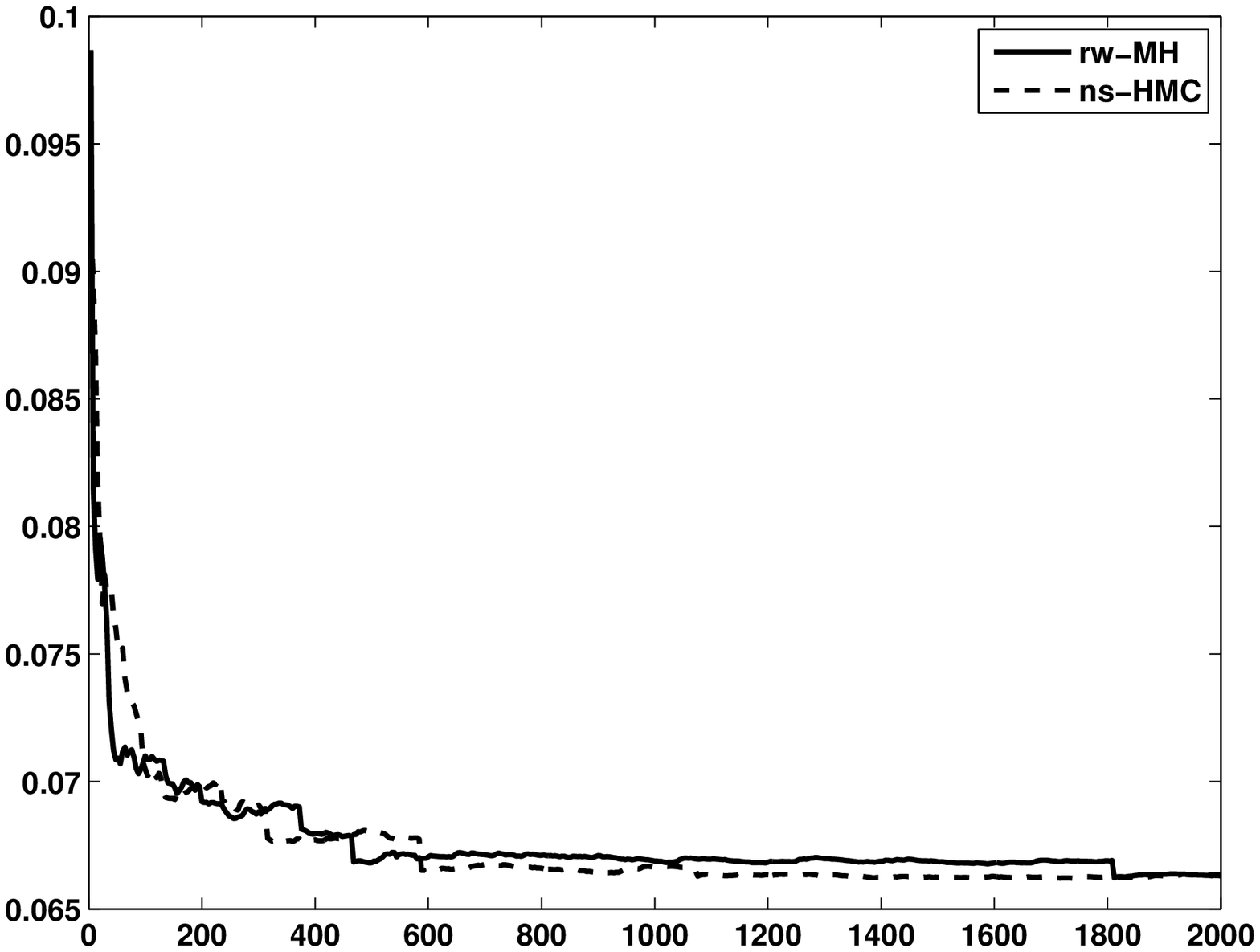}\\
\end{tabular}
 \caption{MSEs between the target 1D GG pdf and the histogram of the generated samples using the rw-MH and ns-HMC algorithms for 
 two different combinations of $p$ and $\lambda$.}\label{fig:samples1d}
 \end{figure}

\begin{figure}[!htp]
\centering
\begin{tabular}{cc}
\rotatebox{90}{\hspace{1.5cm} $p=1$, $\lambda=1$}&\includegraphics[height=4.5cm,width=6.5cm]{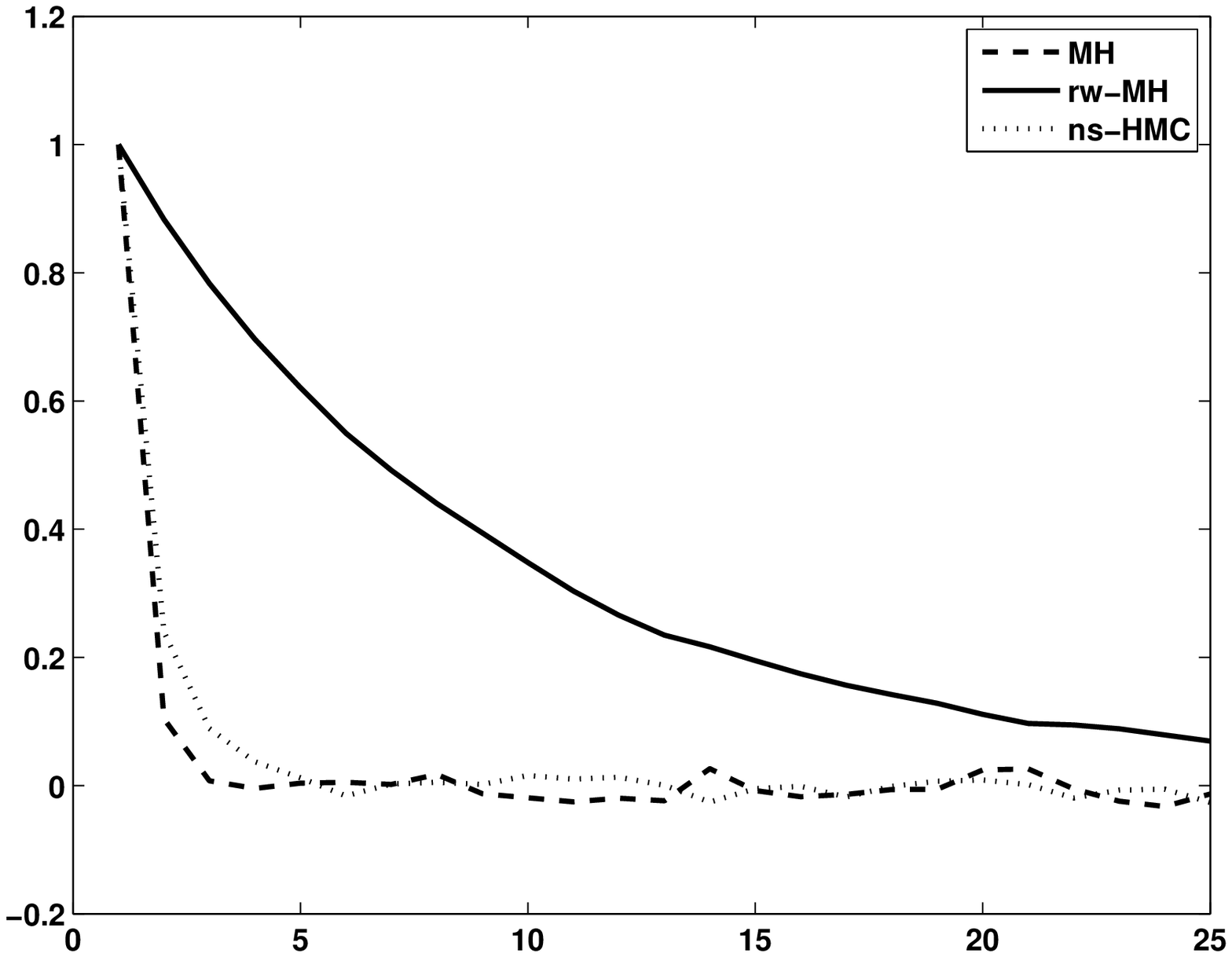}\\
\rotatebox{90}{\hspace{1.5cm}$p=1.5$, $\lambda=1$}&\includegraphics[height=4.5cm,width=6.5cm]{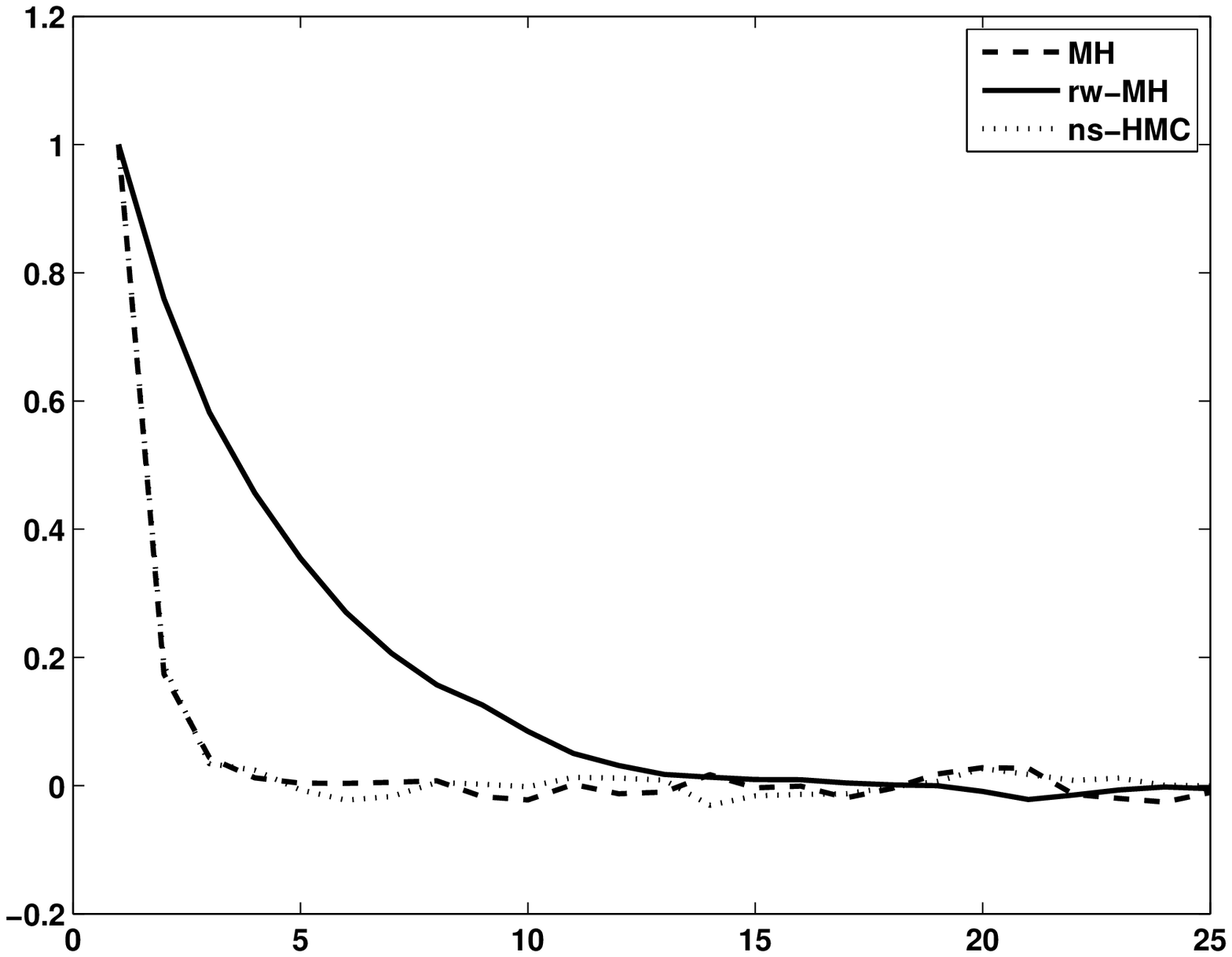}\\
\end{tabular}
 \caption{ACFs of sampled chains using  Metropolis-Hastings (MH) and random walk Metropolis-Hastings (rw-MH) algorithms, in addition to
  the proposed method (ns-HMC) for two values of $(p,\lambda)$.}\label{fig:ACF1d}
 \end{figure}

In the first experiment, a 1D sampling is performed for different values of the shape and scale parameters of a GG distribution 
 ($p$ and $\lambda$). Chains generated using the proposed ns-HMC sampling scheme are compared to the ones 
 obtained with a random walk Metropolis-Hastings (rw-MH) scheme. 
 The rw-MH strategy is used here for comparison since it generally improves the mixing properties of the generated samples when compared 
 to a fixed proposal distribution. 
 Let $x^{(r)}$ be the current sample and $x^{*}$ the proposed one. 
 A Gaussian proposal centered on the current sample with unitary variance is used for the rw-MH algorithm, 
 i.e., $x^{*} \sim \mathcal{N}(x^{(r)},1)$. 
 Fig.~\ref{fig:samples1d} displays the mean square error (MSE) between the target GG pdf and the histogram of the generated samples with respect to 
 the number of sampled coefficients. 
 This figure shows slightly faster convergence for the proposed ns-HMC scheme compared to the rw-MH algorithm. \\
 To further investigate the sampling efficiency, Fig.~\ref{fig:ACF1d} displays the autocorrelation functions (ACFs) of the sampled chains 
 for the same values of $(p,\lambda)$. This figure clearly shows that samples generated using the ns-HMC scheme are less
  correlated than those generated using rw-MH, which corroborates the fast convergence of the ns-HMC scheme. 
  In fact, the proposed technique does not need any adjustment of the proposal variance contrary to the rw-HM algorithm while giving 
  acceptable level of intra-chain correlation. 
  For the sake 
  of comparison, Fig.~\ref{fig:ACF1d} also displays the ACFs of chains sampled using a standard MH algorithm with a  
  centered Gaussian proposal ($x^{*} \sim \mathcal{N}(0,1)$). 
  Indeed, it has been reported that 
  rw-MH increases the correlation level within sampled chains~\cite{Robert_04}, while an MH algorithm 
  provides uncorrelated samples. 
  The comparison between the ACFs corresponding to ns-HMC and MH shows that chains sampled using ns-HMC are as less correlated 
  as the standard MH algorithm with $\mathcal{N}(0,1)$ proposal.

%

\subsection{Experiment 2: multivariate sampling}
\begin{figure*}[!htp]
\centering
\begin{tabular}{c|cccc}
\raisebox{2.2cm}{2D}&\rotatebox{90}{\hspace{2cm}MSE}&\includegraphics[height=5.5cm,width=7.5cm]{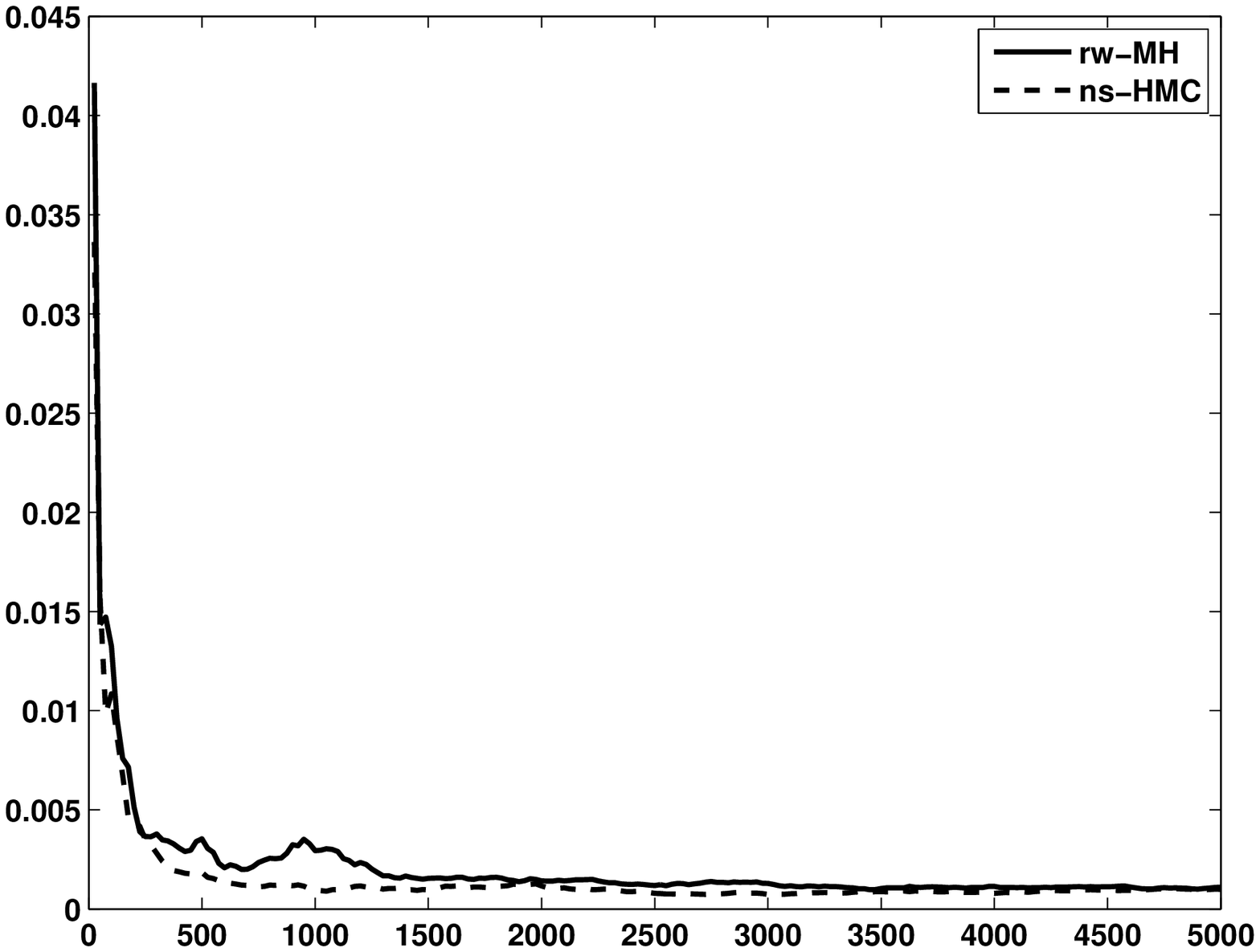}&
\rotatebox{90}{\hspace{2cm}MSE}&\includegraphics[height=5.5cm,width=7.5cm]{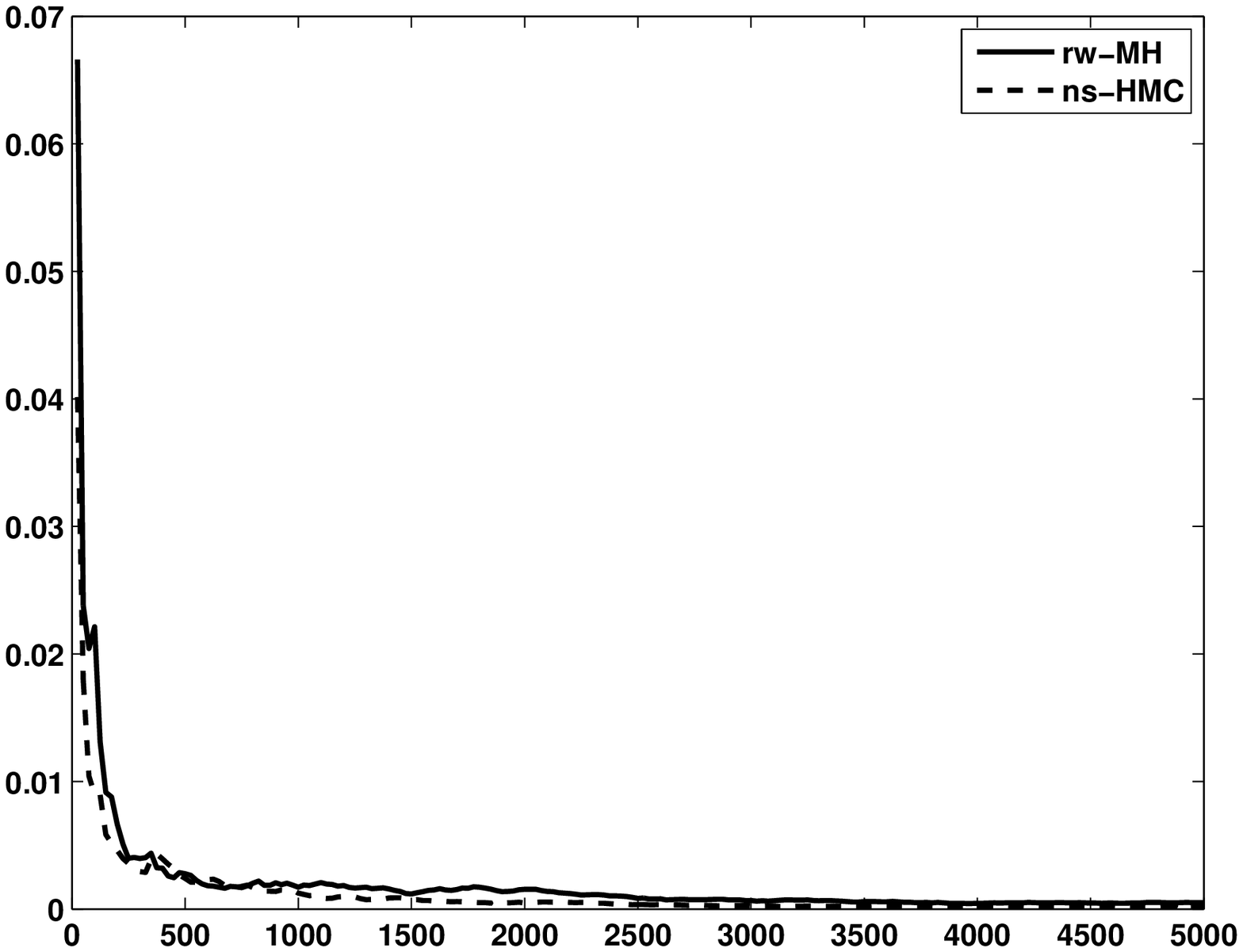}\\
&&$p=1$, $\lambda=1$&&$p=1.5$, $\lambda=1$\\
\raisebox{2.2cm}{3D}&\rotatebox{90}{\hspace{2cm}MSE}&\includegraphics[height=5.5cm,width=7.5cm]{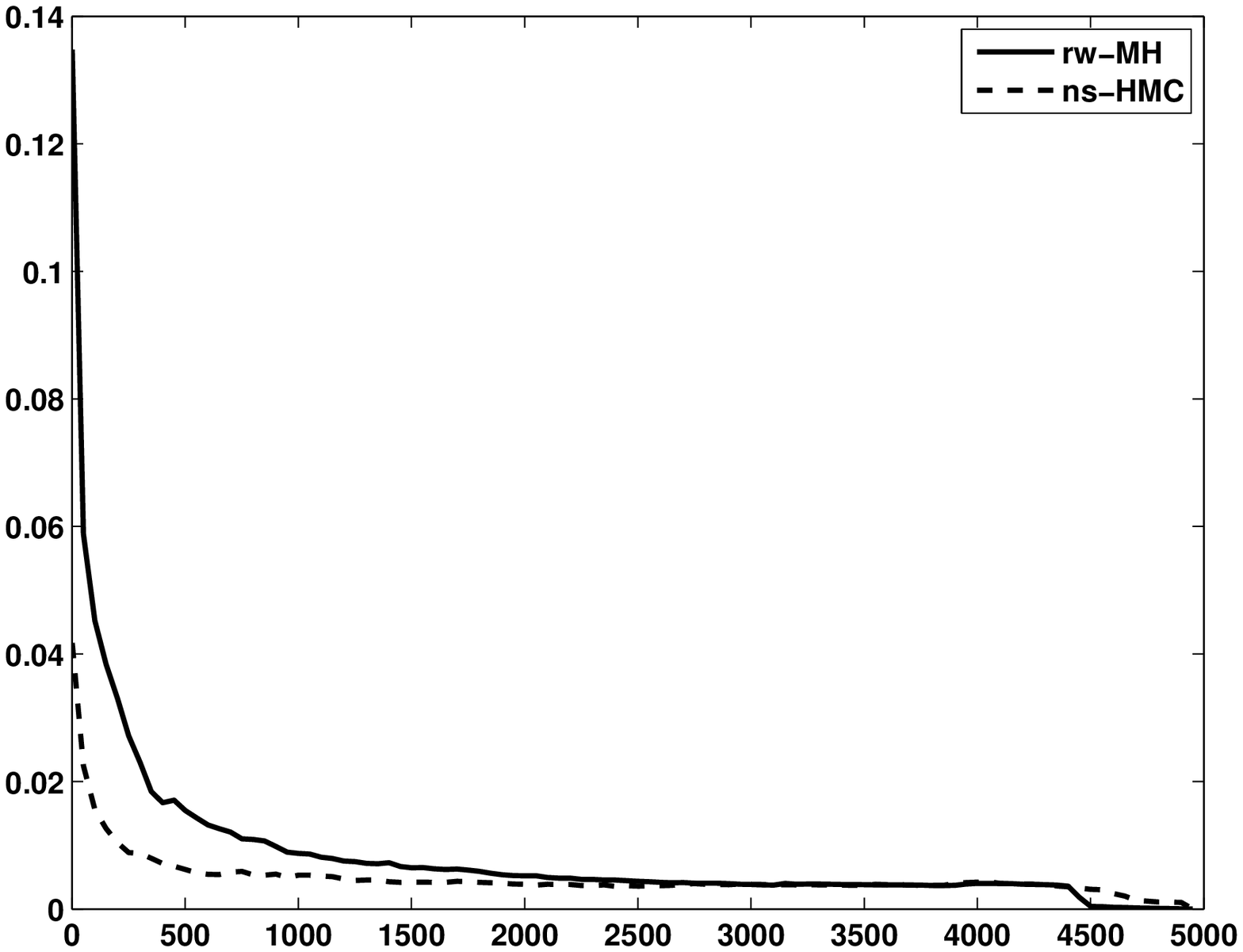}&
\rotatebox{90}{\hspace{2cm}MSE}&\includegraphics[height=5.5cm,width=7.5cm]{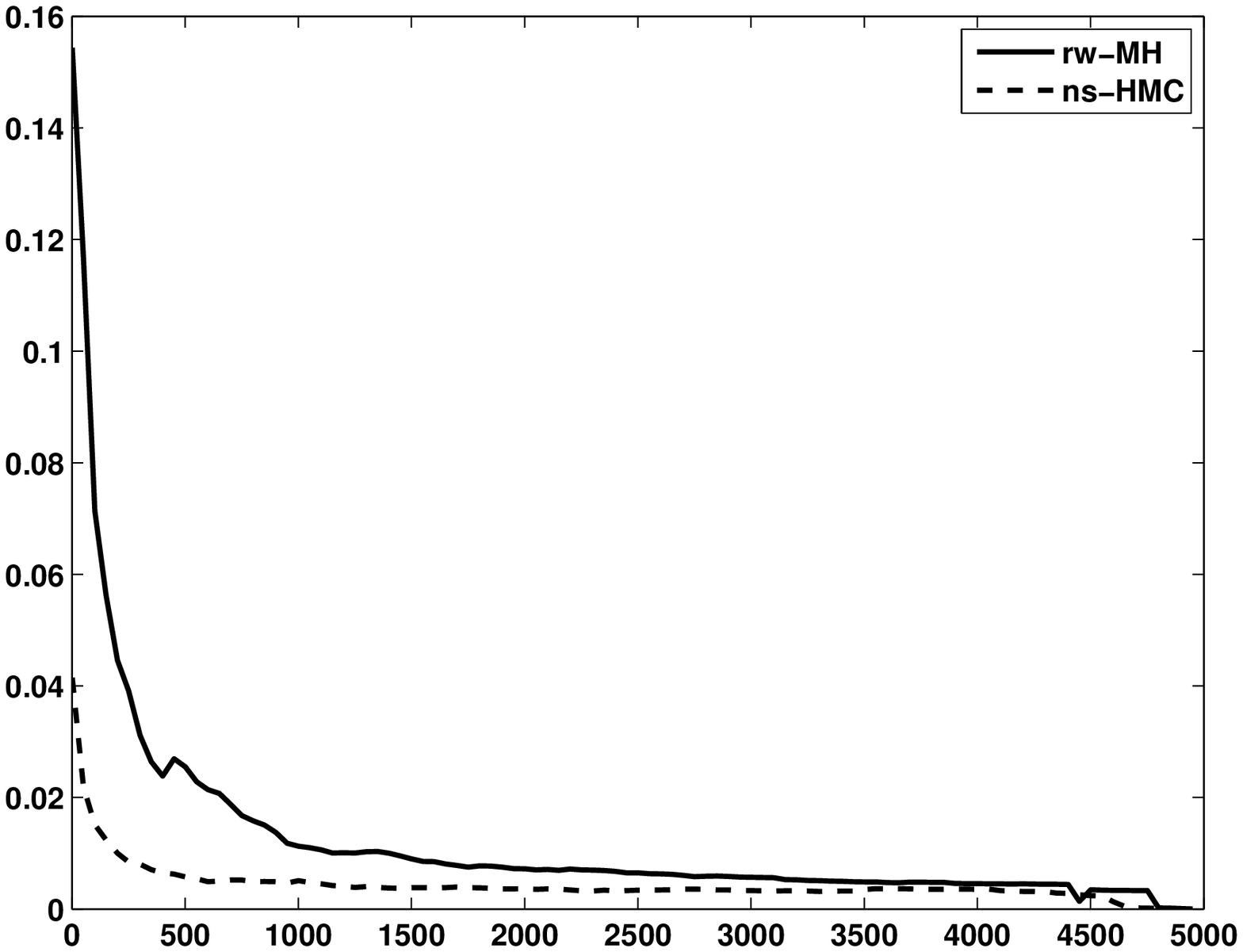}\\
&&$p=1$, $\lambda=1$&&$p=1.5$, $\lambda=1$\\
\raisebox{2.2cm}{4D}&\rotatebox{90}{\hspace{2cm}MSE}&\includegraphics[height=5.5cm,width=7.5cm]{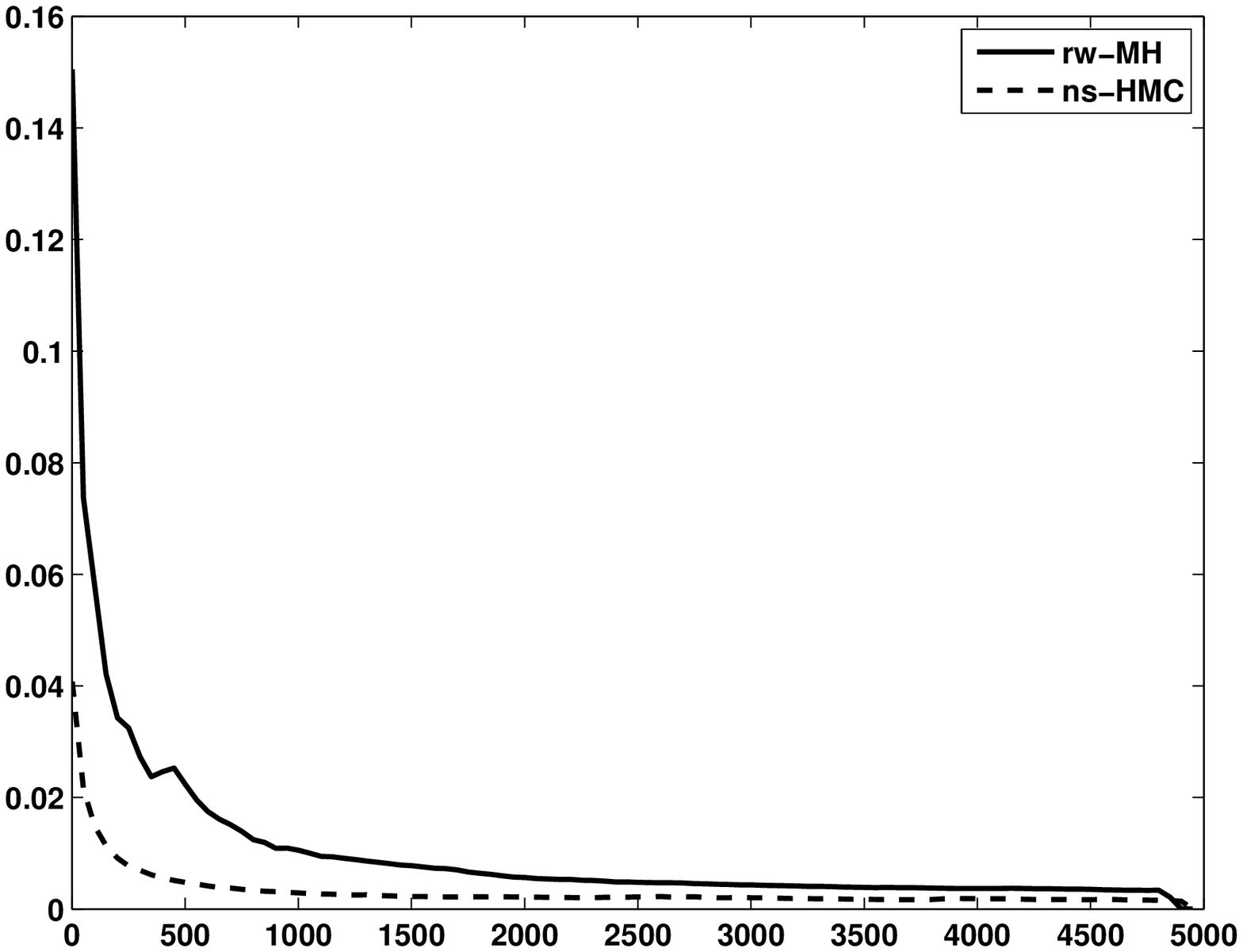}&
\rotatebox{90}{\hspace{2cm}MSE}&\includegraphics[height=5.5cm,width=7.5cm]{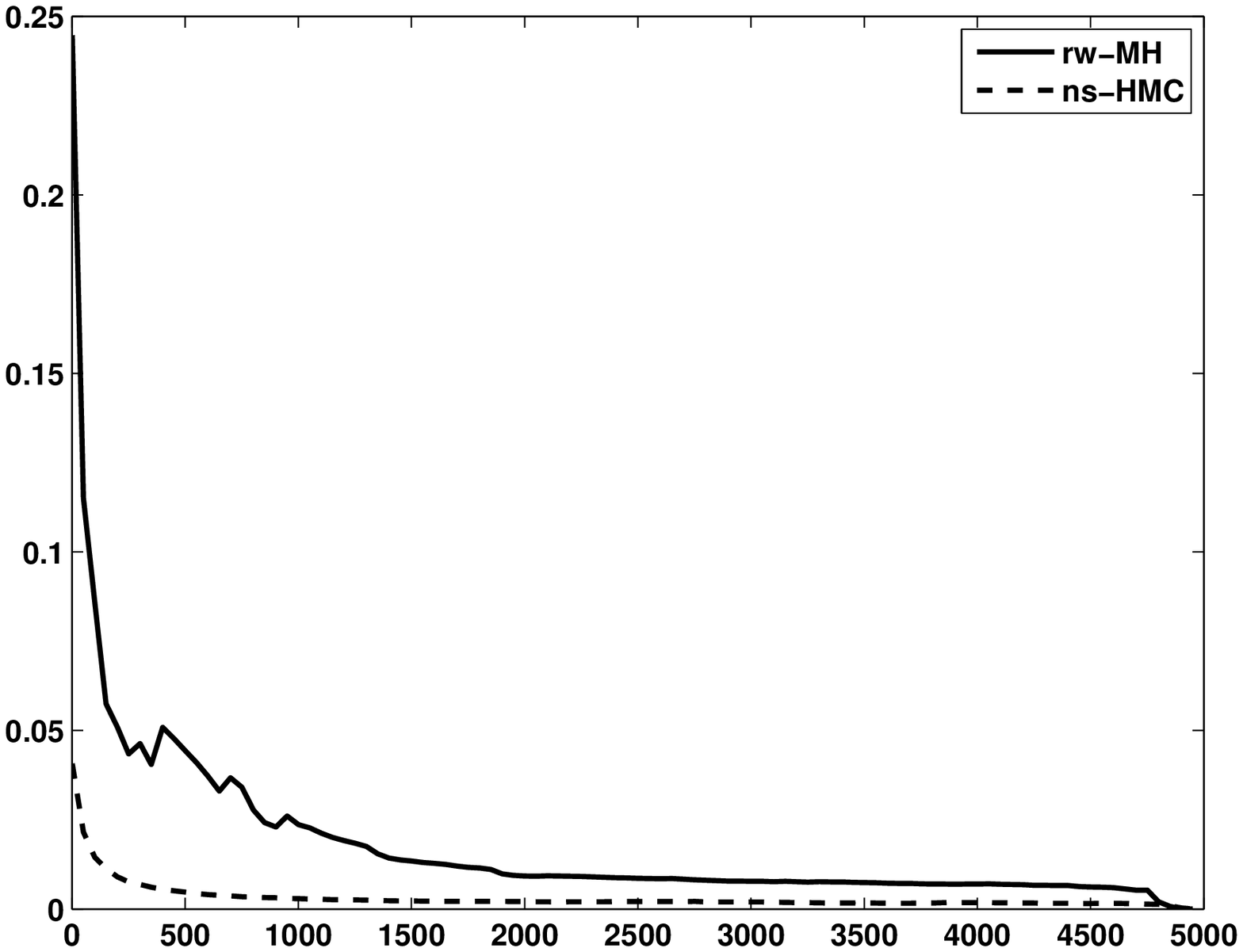}\\
&&$p=1$, $\lambda=1$&&$p=1.5$, $\lambda=1$\\
\end{tabular}
 \caption{MSEs between the target GG pdf and the histogram of the generated samples for the rw-MH and ns-HMC algorithms.
 \label{fig:mse1}}
 \end{figure*}

In this experiment, sampling is performed according to a multivariate GG distribution. 
First, sampling using rw-MH is performed with a large number of iterations (long burn-in period) 
so that the target distribution is guaranteed to be reached. The histogram of the obtained samples is calculated 
and corresponds to our ground truth. 
Then, samplings using the proposed ns-HMC method and the rw-MH scheme are performed on the same dataset. 
The MSE between the obtained histograms and the ground truth are computed versus the number of iterations of the sampler. 
Fig.~\ref{fig:mse1} displays this MSE for the multivariate GG sampling and for different values of the 
scale and shape parameters ($\lambda$ and $p$). Note that simulations have been performed for the 2D, 3D and 4D cases. 
The reported MSE values show that the target distribution is more rapidly approached 
using ns-HMC compared to rw-MH. Specifically, for the 2D case, 
the ns-HMC scheme converges after about 500 iterations for $p=1$ (resp. 1000 iterations for $p=1.5$), while the rw-MH 
sampling needs about 3500 iterations for $p=1$ and $p=2$. When looking at the same curves in Fig.~\ref{fig:mse1} for the 3D and 4D cases, 
it is worth noticing that the gap between the two methods in terms of convergence speed increases with the problem dimensionality. 
This effect is confirmed for the shape and scale parameters. This corroborates the usefulness of the proposed 
ns-HMC scheme especially for large data sampling where the convergence speed of the standard MH or rw-MH algorithms is altered by the 
size of the data. 

\subsection{Experiment 3: denoising}
\label{subsec:exp3}
\begin{figure}[!htp]
\centering

\begin{tabular}{cc}
 \footnotesize{Reference}&
\begin{minipage}{4cm}
\centering
 \footnotesize{Noisy\\
 SNR = 5.68~dB, 
 SSIM = 0.699}
\end{minipage}
\\
\includegraphics[height=4cm,width=4cm]{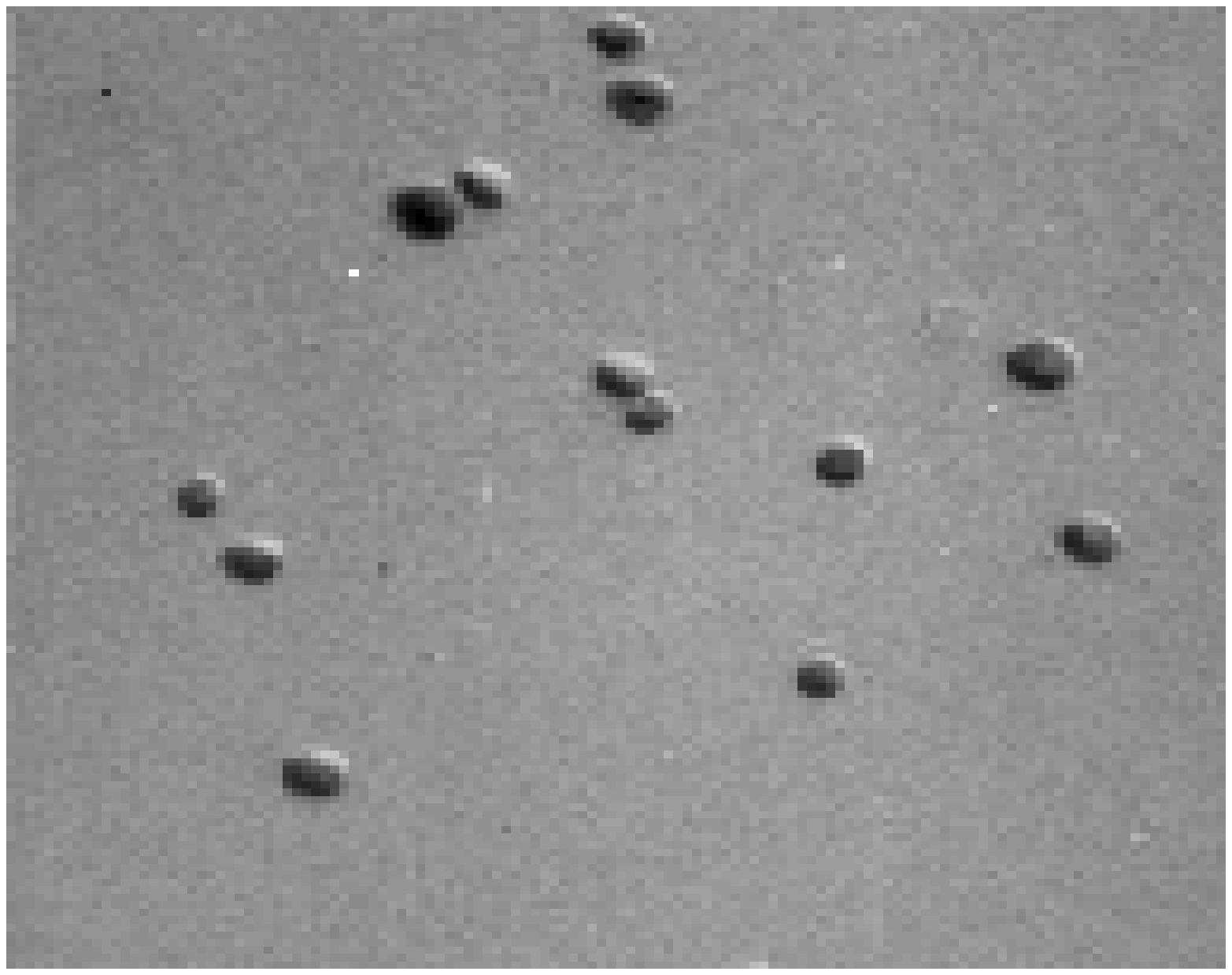}&
\includegraphics[height=4cm,width=4cm]{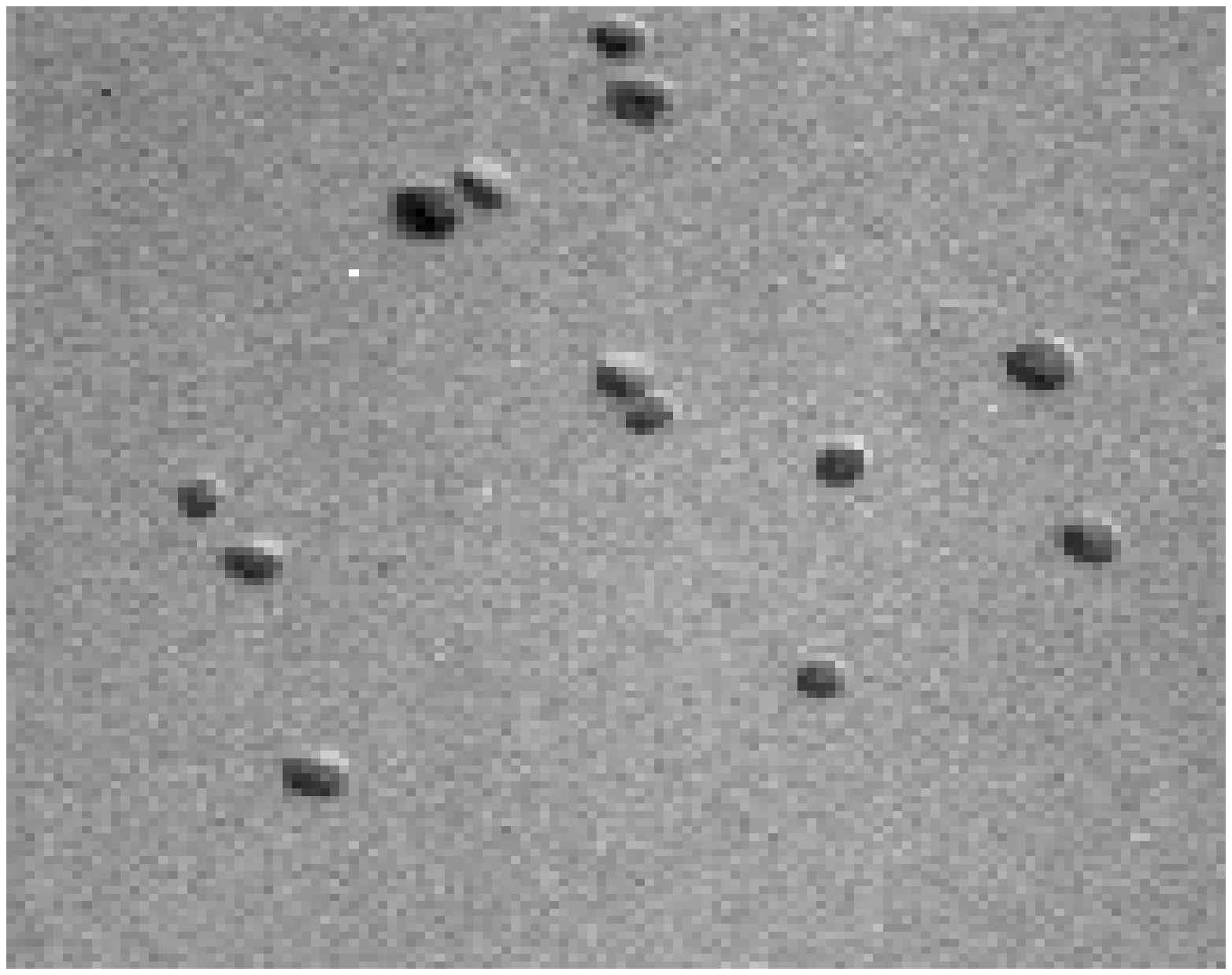}\\
 \includegraphics[height=4cm,width=4cm]{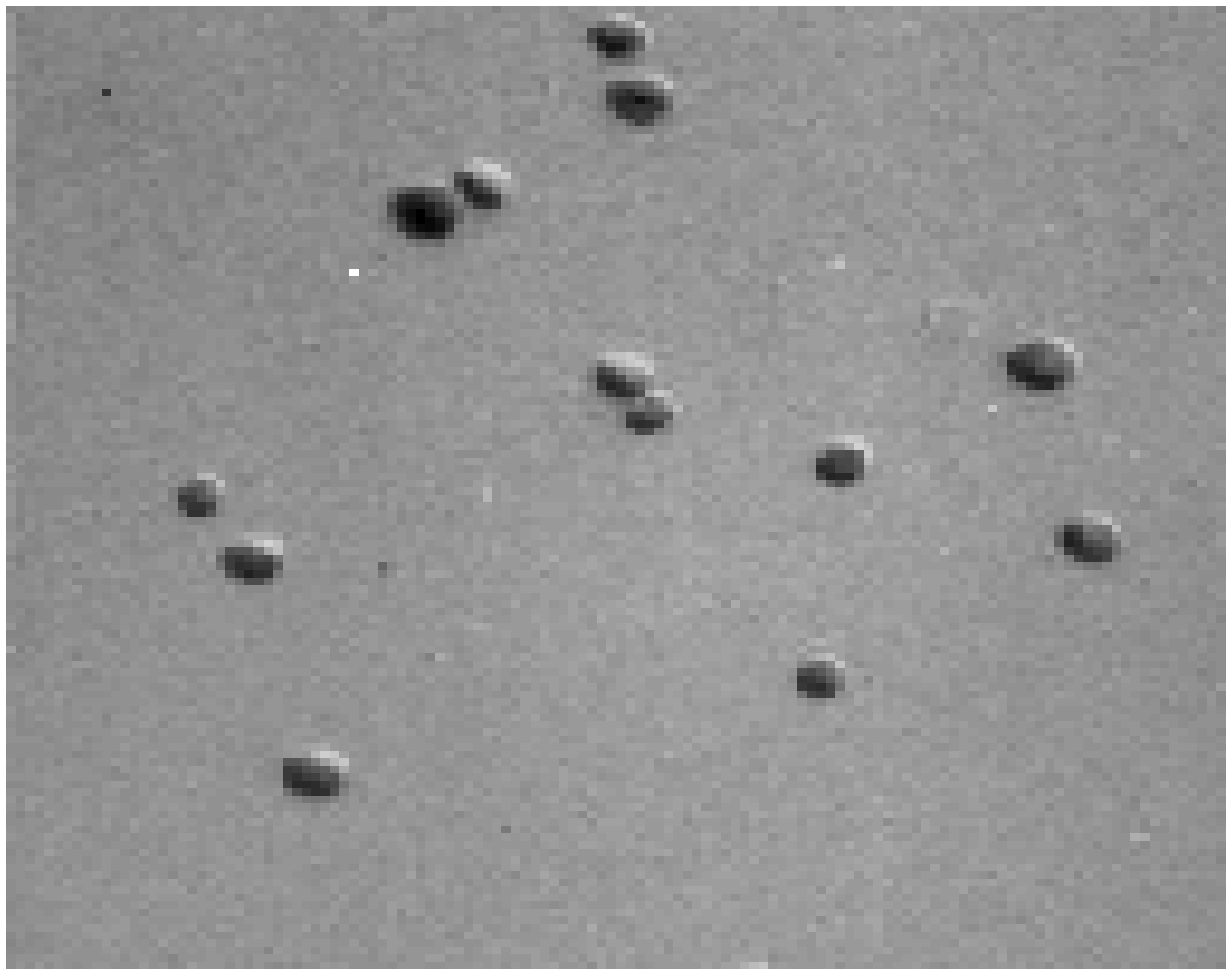}&
\includegraphics[height=4cm,width=4cm]{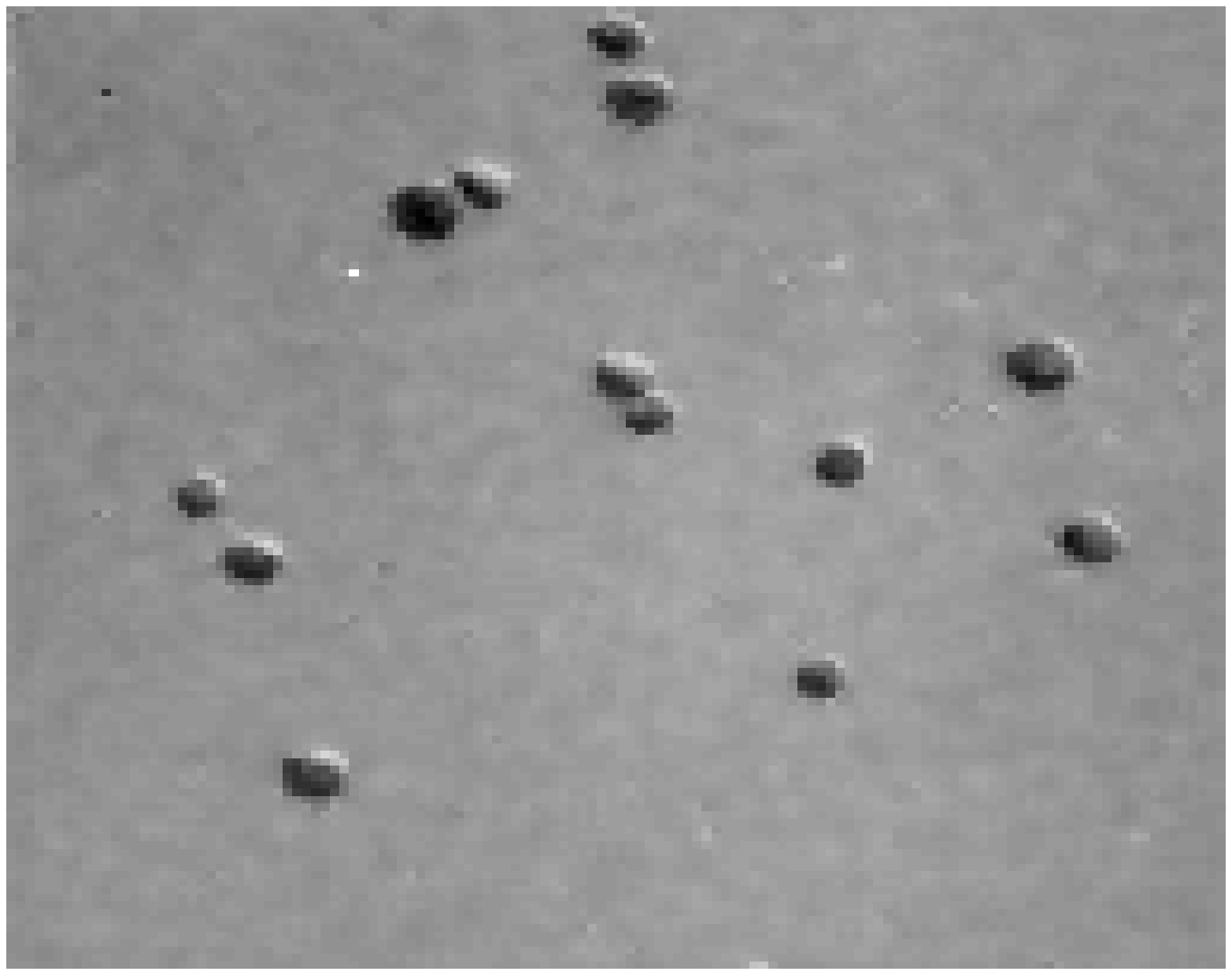}\\
\begin{minipage}{4cm}
\centering
 \footnotesize{Algorithm~\ref{algo:denoising}\\
 SNR = 20.48~dB, 
 SSIM = 0.985}
\end{minipage}
& \begin{minipage}{4cm}
\centering
 \footnotesize{Wiener\\
 SNR = 8.44~dB, 
 SSIM = 0.764}
\end{minipage} 

\end{tabular}
\caption{Reference (top-left), noisy (top-right) and denoised images using Algorithm~\ref{algo:denoising} (bottom-left) and the 
Wiener filter (bottom-right).
\label{fig:denoising}}
\end{figure}

In this experiment, the performance of the proposed ns-HMC sampling algorithm is analyzed for an image denoising
problem.  
The 2D image of size $N = 128 \times 128$ displayed in Fig.~\ref{fig:denoising}[top-left] has been used as a ground truth for this example. 
An independent identically distributed additive Gaussian noise of variance $\sigma_n^2=40$ has been added to this image to obtain 
the noisy image depicted in Fig.~\ref{fig:denoising}[top-right]. 
The objective of this third experiment is to promote the sparsity of the wavelet coefficients associated with the target image. 
To this end, we express the image formation model as function of the wavelet coefficients $\xb \in \RR^{N}$ which
are related to the ground truth image $\zb$ through the relation $\zb = F^{-1} \xb$ where $F^{-1}\in \RR^{N \times N}$ denotes
the dual frame operator. The analysis frame operator thus corresponds to $F\in \RR^{N \times N}$ and as orthonormal bases are considered here,
the dual frame operator reduces to the inverse operator yielding $F^{-1}F=FF^{-1}=\Id$. The observation model can thus be expressed as
\begin{equation}
 \yb = F^{-1} \xb + \nb
\end{equation}
where $\yb \in \RR^{N}$ is the observed image, $\xb \in \RR^{N}$ contains the unknown wavelet coefficients and $\nb \in \RR^{N}$ is
the additive noise. Note that the denoised image $\widehat{\zb}$ can be easily recovered from the estimated wavelet coefficients $\widehat{\xb}$ by taking $\widehat{\zb} = F^{-1} \widehat{\xb}$.\\
Based on this model and the Gaussian likelihood assumption, a hierarchical Bayesian model has been built using an independent Laplace prior 
for the wavelet coefficients \cite{tipping01,seeger08}
\begin{equation}
 f(\xb;\lambda) = \left(\frac{1}{2 \lambda}\right)^{N} \exp{\left(-\frac{||\xb||_1}{\lambda}\right)}
\end{equation}
where $\lambda$ is an unknown parameter that is estimated within the proposed Bayesian algorithm. 
More precisely, an inverse gamma prior distribution is assigned to $\lambda$ \cite{Chaari_eusipco_2013,dobigeon_09}
\begin{equation}
 f(\lambda| a,b) = \mathcal{IG}(\lambda|a,b)  = \frac{b^a}{\Gamma(a)} \lambda^{- a -1} 
 \exp{\left(-\frac{b}{\lambda}\right)}
\end{equation}
where $\Gamma(.)$ is the gamma function, and $a$ and $b$ are fixed hyperparameters 
(in our experiments these hyperparameters have been set to $a = b = 10^{-3}$).\\
Using a Jeffrey's prior for the noise variance ($\sigma_n^2 \sim \frac{1}{\sigma_n^2} 1_{\RR^+}(\sigma_n^2)$), the full 
posterior of this denoising model can be derived. The associated Gibbs sampler generates samples according to the conditional 
distributions of the posterior. The conditional distribution of the wavelet coefficients $\xb$ writes
\begin{equation}
\label{eq:postx}
 f(\xb|\yb,\sigma_n^2,\lambda) \propto \exp{\left[-U(\xb)\right]}
\end{equation}
where the energy function $U$ is defined by $U(\xb) = \frac{||\xb||_1}{\lambda} + \frac{|| \yb - F^{-1}\xb||^2_2}{2 \sigma_n^2 }$.
Sampling according to this distribution is performed using the proposed ns-HMC scheme, which requires the calculation of the proximity operator 
of its energy function given by
\begin{align}
\label{eq:prox}
 \prox_{U}(\xb) = & \prox_{||\cdot||_1/(1+\alpha)} \left( \frac{\xb + F\yb}{1+\alpha} \right)
\end{align}
where $\alpha = \dfrac{1}{\sigma_n^2}$ and $\prox_{||\cdot||_1/(1+\alpha)}$ can easily be calculated using standard properties of the proximity 
operators~\cite{Chaux_C_07,Combettes_PL_08,Bauschke_combettes2011}. See Appendix~\ref{append:prox} for more details.\\
Regarding the noise variance and the prior hyperparameter, straightforward calculations lead to the following 
conditional distributions which are 
easy to sample 
\begin{align}
 \sigma_n^2|\xb,\yb &\sim \mathcal{IG}\Big(\sigma_n^2|N/2,||\yb - F^{-1}\xb||^2/2\Big)\label{eq:post_sigma2} \\
 \lambda|\xb,a,b &\sim \mathcal{IG}\Big(\lambda|a + N,b + ||\xb||_1\Big)\label{eq:post_lambda}
\end{align}
where $\mathcal{IG}$ is the inverse gamma distribution. 
The estimation of the denoised image is performed based on the sampled wavelet coefficients after an appropriate 
burn-in perior, i.e., after convergence of the Gibbs sampler in 
Algorithm~\ref{algo:denoising}.

\begin{algorithm}\label{algo:denoising}
\SetAlgoLined
- Initialize with some $\vect{x}^{(0)}$.\\
\For{$r=1,2\ldots$}{
- Sample ${\sigma_n^2}^{(r)}$ according to \eqref{eq:post_sigma2}\;
- Sample $\lambda^{(r)}$ according to \eqref{eq:post_lambda}\;
- Sample $\xb^{(r)}$ according to its conditional distribution using the proposed ns-HMC scheme\;
}
- After convergence, compute the MMSE estimator $\widehat{\xb}$ and return the 
estimated image $\widehat{\zb} = F^{-1}\widehat{\xb}$.
\caption{Gibbs sampler for image denoising.}
\end{algorithm}

An example of denoised image using Algorithm~\ref{algo:denoising} is displayed in Fig.~\ref{fig:denoising}[bottom-left]. 
For the sake of comparison, a denoised image using the Wiener filter is displayed in Fig.~\ref{fig:denoising}[bottom-right]. From a visual point of view, 
we can easily notice that Algorithm~\ref{algo:denoising} provides a better denoised image compared to the Wiener filter. 
Quantitatively speaking, the evaluation of the noisy and 
denoised images is based on both SNR (signal to noise ratio) and SSIM~\cite{Wang_04} (structural similarity). These values are directly 
reported in the figure and show the efficiency of the denoising algorithm based on the proposed ns-HMC technique to sample from the 
conditional distribution of the wavelet coefficients $\xb$. 
As regards the computational time, only 1000 iterations are 
necessary for the proposed 
algorithm involving a burn-in period of 500 iterations, taking around 9 seconds on a 64-bit 2.00GHz i7-3667U architecture with a Matlab implementation. For the ns-HMC step, the second scheme has been used 
with $L_f=10$.

\section{Conclusion}
\label{sec:cl}
This paper proposed a solution to make feasible the use of Hamiltonian dynamics for sampling according to log-concave probability 
distributions with non-smooth energy functions. The proposed sampling technique relies on some interesting results from 
convex optimization and Hamiltonian Monte Carlo methods. 
More precisely, proximity operators were investigated to address the non-differentiability problem of the energy function related to the 
target distribution. Validation results showed that the proposed technique provides faster convergence and interesting decorrelation properties 
for the sampled chains
 when compared to more standard methods such as the random walk Metropolis Hastings algorithm. 
 The proposed technique was evaluated on synthetic data and 
applied to an image denoising problem. 
Our results showed that the use of proximity operators in a Hamiltonian Monte Carlo method allows faster convergence to the target 
distribution to be obtained. This conclusion is particularly important for large scale data sampling since the 
 gain in convergence speed increases with the problem dimensionality. 
 In a future work, we will focus on the investigation of this technique for sparse signal recovery where a non-tight linear operator 
 is involved in the observation model.

%

\appendix

\subsection{Volume preservation} \label{append:volume} \hfill \\
This appendix studies the volume preservation condition both for 
the Hamiltonian dynamics and the proposed discretization 
schemes. Volume preservation is a key 
property of sampling algorithms involving an acceptance-rejection step 
such as the Metropolis within Gibbs algorithm since it allows simpler acceptance probability to be obtained. 
\subsubsection{Volume preservation for the Hamiltonian dynamics}\hfill \\
Akin to \cite{neal2010mcmc}, we consider here volume preservation for Hamiltonian dynamics for the one dimensional case. 
The multi-dimensional case can then be handled through simple generalizations. 
Let us denote by $\mathcal{F}_\delta$ (see Section~\ref{sec:HMCmethods}) the mapping between the state at time $t$, denoted 
by $(x(t),q(t))$, and the state $(x(t+\delta),q(t+\delta))$ at time $t+\delta$. 
For $\delta$ small enough, $\mathcal{F}_\delta$ can be approximated by~\cite{neal2010mcmc}
\begin{equation}\label{eq:Hamiltonianb}
 \mathcal{F}_\delta(q,x) = 
 \left[
\begin{array}{c}
q\\
x\\
\end{array}
\right] 
+
\delta \left[
\begin{array}{c}
dq/dt\\
dx/dt\\
\end{array}
\right] 
+ O(\delta^2)
\end{equation}
where $O(\delta^2)$ involves terms of order $\delta^2$ or higher. 
After replacing the time derivatives of \eqref{eq:hammotion} in \eqref{eq:Hamiltonianb}, and accounting for the fact that the Hamiltonian may be non-differentiable 
with respect to $x$, the generalized Jacobian matrix can be written as

\begin{align}
 \mathcal{J}_\delta = &\left[
\begin{array}{cc}
1+\delta \dfrac{\partial^2 H_\thetab}{\partial  q \partial x}&\delta \dfrac{\partial^2 H_\thetab}{\partial q^2}\\
-\delta \dfrac{\partial^2 H_\thetab}{\partial  q ^2}& 1-\delta \dfrac{\partial^2 H_\thetab}{ \partial x\partial  q}\\
\end{array}
\right] 
+O(\delta^2)
\end{align}
where $\partial$ denotes the sub-gradient and $\dfrac{\partial^2 H_\thetab}{\partial  q \partial x}$ is an element of 
the second-order sub differential with respect to $q$ and $x$. 
The determinant of this matrix can therefore be written as
\begin{align}
\mathrm{det}(\mathcal{J}_\delta) &=1+\delta \dfrac{\partial^2 H_\thetab}{\partial  q \partial x} - \delta \dfrac{\partial^2 H_\thetab}{ \partial x\partial  q} 
 +  O(\delta^2) \nonumber \\
 &=1+ O(\delta^2)
\end{align} where $\mathrm{det}(A)$ is the determinant of the matrix $A$.
Following the construction proposed in \cite{neal2010mcmc}, 
it turns out that for some time interval $s$ that is not close to zero, $\mathrm{det}(\mathcal{J}_s)=1$, which means that the 
transformation $\mathcal{F}_s$  ensures volume preservation. \\
\subsubsection{Volume preservation for the proposed discretization schemes}\hfill \\
It is woth noticing here that the two modified leapfrog discretization schemes $T'_{s}$ and $T''_{s}$ defined respectively in 
\eqref{eq:leapfrog1bb}-\eqref{eq:leapfrog3bb} and   \eqref{eq:leapfrog1b}-\eqref{eq:leapfrog3b}, 
as the original leapfrog scheme defined in \eqref{eq:leapfrog1}-\eqref{eq:leapfrog3}, preserve volume since they are shear 
transformations. The interested reader can refer to~\cite{neal2010mcmc} or \cite[page 121]{Brooks_11} for more details.

\subsection{Proximity operator calculation for the experiment of Section~\ref{subsec:exp3}} \label{append:prox} \hfill \\
The energy function considered in this appendix is the one involved in the conditional distribution of the wavelet coefficients 
in~\eqref{eq:postx}, i.e.,
\begin{equation}
 U(\xb) = \dfrac{\alpha}{2} || \yb - F^{-1}\xb||^2_2 + \varphi{(\xb)}
\end{equation}
where $\alpha = 1/\sigma^2_n$ and $\varphi(\xb) = \frac{||\xb||_1}{\lambda}$. 
In order to use the proposed ns-HMC sampling algorithm, the proximity operator of the function $U$ has to be calculated. 
Following the standard definition of the proximity operator~\cite{Moreau_65,Chaux_C_07}, we can write

\begin{align}
 \prox_U(\xb) = \pb \Leftrightarrow&  \xb - \pb \in \partial U(\pb) \nonumber \\
 \Leftrightarrow & \xb - \pb \in \partial \varphi(\pb) + \alpha \pb - \alpha F \yb \nonumber \\
  \Leftrightarrow & \xb + \alpha F \yb - (\alpha+1) \pb \in  \partial \varphi(\pb) \nonumber \\
 \Leftrightarrow & \dfrac{\xb + \alpha F \yb}{\alpha+1} - \pb \in \partial \varphi/(\alpha+1)(\pb) \nonumber \\
 \Leftrightarrow &  \pb = \prox_{\varphi/(\alpha+1)} \left( \dfrac{\xb + F\yb}{\alpha+1} \right)
\end{align}
which proves the expresssion of the proximity operator given in~\eqref{eq:prox}.

\bibliographystyle{IEEEbib}
\bibliography{biblio_hamprox}
\end{document}